\newcommand{\EQ}[1] {Equation~(\ref{#1})}
\newcommand{\SEC}[1] {Section~\ref{#1}}

\newcommand{\FIG}[1] {Figure~\ref{#1}}
\newcommand{\TAB}[1] {Table~\ref{#1}}

\newcommand{\VEC}[1] {{\boldsymbol{{ #1}}}}
\newcommand{\PPD}{$P-\dot{P}$ } 

\newcommand{\FTH}{$F_{1000}$}

\documentclass[useAMS,usenatbib, twocolumn]{mn2e}
\usepackage{graphicx, amsmath}

\title[Application of the Gaussian mixture model]
{Application of the Gaussian mixture model in pulsar astronomy --
pulsar classification and candidates ranking for {\it Fermi} 2FGL catalog}

\author[K.~J.~Lee et al.]{
K.~J.~Lee $^{1}$\thanks{Email: kjlee@mpifr-bonn.mpg.de},
L.~Guillemot ${^1}$,
Y.~L.~Yue $^{3}$,
M.~Kramer $^{1,2}$,
D.~J.~Champion $^{1}$
\\
$^1${Max-Planck-Institut f\"ur Radioastronomie, Auf dem H\"ugel 69,
	D-53121 Bonn, Germany} \\
$^2${Jodrell Bank Centre for Astrophysics, University of Manchester,
	Manchester M13 9PL, UK} \\
	$^3${ National Astronomical Observatories, Chinese Academy of Sciences,
A20 Datun Road, Chaoyang District, Beijing 100012, China.
	}\\
	}
\begin{document}

\date{\today} \pagerange{\pageref{firstpage}--\pageref{lastpage}}
\pubyear{2010} \maketitle \label{firstpage}

\begin{abstract} Machine learning, algorithms to extract empirical
knowledge from data, can be used to classify data, which is one of the
most common tasks in observational astronomy. In this paper, we focus
on Bayesian data classification algorithms using the Gaussian mixture
model and show two applications in pulsar astronomy.  After reviewing the
Gaussian mixture model and the related Expectation-Maximization algorithm,
we present a data classification method using the Neyman-Pearson test. To
demonstrate the method, we apply the algorithm to two classification
problems.  Firstly, it is applied to the well known period-period
derivative diagram, where we find that the pulsar distribution can be
modeled with six Gaussian clusters, with two clusters for millisecond
pulsars (recycled pulsars) and the rest for normal pulsars. From this
distribution, we derive an empirical definition for millisecond pulsars
as $\frac{\dot{P}}{10^{-17}}  \leq3.23\left( \frac{P}{100\, \textrm{ 
ms}}\right)^{-2.34}$. The two millisecond pulsar clusters may have
different evolutionary origins, since the companion stars to these pulsars in 
the two
clusters show different chemical composition. Four clusters are found for 
normal pulsars.
Possible implications for these clusters are also discussed.
Our second example is to calculate the likelihood of unidentified
\textit{Fermi} point sources being pulsars and rank them accordingly. In
the ranked point source list, the top 5\% sources contain 50\% known
pulsars, the top 50\% contain 99\% known pulsars, and no known active
galaxy (the other major population) appears in the top 6\%. Such a ranked
list can be used to help the future follow-up observations for finding
pulsars in unidentified \textit{Fermi} point sources.  \end{abstract}

\section{Introduction} 

A common and important task in observational astronomy is to find or
select a certain type of objects from a sample of candidates according
to particular physical properties. Examples include selecting active
galaxy candidates from a multi-color optical photometry survey,
selecting good pulsar candidates from a large number of candidates
produced in pulsar searches. Such tasks are time consuming,
especially when the number of candidates is large. In this situation,
computers can offer significant help when the selection rules can be
derived based on prior experiences or physical considerations. However,
for some applications, it is hard to determine the \emph{a priori}
criteria and one has to search for the selection criterion using the
\emph{empirical} knowledge embedded in the data themselves.  For example,
different types of sources usually form clusters in different regions
of parameter space. When a large population of candidates is available,
the clustering becomes statistically significant, and one can then use
this to determine the selection criterion.

In order to build up the empirical selection criterion, we need to find
a method to \emph{learn} the knowledge from the data and apply this to
generate the selection rules. Machine learning algorithms are designed
to extract empirical knowledge from a sample of data, and improve its
performance based on the knowledge it learnt.  Machine learning algorithms
also contain methods to classify data. There are already many successful
applications of machine learning algorithms since the 1960s, some of
which were recently used in the pulsar community (e.g.  \citealt{EMK11}).
We refer to \cite{Th09} for the details of such algorithms and their
applications in broader fields. Clearly, machine learning and related
classification algorithms can help to determine the criteria required
to select desirable objects from a large sample of candidates in the
context of observational astronomy.

This paper demonstrates the application of Gaussian mixture
model (GMM) in the context of pulsar astronomy. The GMM, which we use here,
is one type of un-supervised learning algorithms based on Bayesian
decision theory \citep{NR3}.  It assumes that the data clusters in
parameter space follow a superposition of several multivariate Gaussian
distributions. The parameters of each cluster are determined from
data using the Expectation-Maximization (EM) algorithm.  We use two
examples to show the application of GMM in pulsar astronomy. Our first
example is to classify pulsars in the parameter space of pulsar period
($P$) and period derivative ($\dot{P}$), and the second example is to
calculate the likelihood of a gamma-ray point source being a pulsar. Here
gamma-ray point source parameters are from the \textit{Fermi} gamma-ray
Space Telescope Large Area Telescope 2-year Point Source Catalog (2FGL catalog,
\citealt{LAT2FGL}).

This article is organized as follows. We introduce the statistical
technique, the GMM, in \SEC{sec:stmt}. We use two problems as examples
to show properties and applications of GMM in \SEC{sec:app}. The first
application in \SEC{sec:psrapp} is to find an empirical definition for
millisecond pulsars (MSPs) from the period-period derivative ($P-\dot{P}$)
distribution of known pulsars. The second application in \SEC{sec:ferapp}
is to describe the 2FGL catalog point source distribution, to calculate
the likelihood of a particular source being a pulsar, and to produce a
pulsar candidate list for later confirmation observations. Conclusions
and discussions are given in \SEC{sec:con}.

\section{Gaussian Mixture Model And Likelihood Ranking} \label{sec:stmt}

In this section, we introduce the basic concepts of GMM as well as
related techniques to classify the data in parameter space.

The GMM is a probabilistic model to describe the distribution of data
with clusters in the parameter space, where each cluster is assumed
to follow the Gaussian distribution. For a total of $m$ clusters in a
$n$-dimensional parameter space, the probability distribution $P(\VEC{x})$
of data $\VEC{x}$  is given by a weighted summation of all $m$ Gaussian 
clusters,
i.e.
\begin{equation} P(\VEC{x})=\sum_{k=1}^{m} P_{k} P(\VEC{x}| \VEC{\mu}_{k},
	\VEC{\Sigma}_{k})\,, \label{eq:e1} \end{equation}
	where the mixture weight of the $k$-th Gaussian is $P_{k}$ and the
	distribution of each individual Gaussian cluster is
\begin{equation} P(\VEC{x}| \VEC{\mu}_{k},
		\VEC{\Sigma}_{k}) =
		\frac{\exp\left[-\frac{1}{2} (\VEC{x-\mu}_{k})\cdot
		\VEC{\Sigma}_{k}^{-1}\cdot(\VEC{x-\mu}_{k})\right]}{(2\pi)^{n/2}
		\sqrt{|\VEC{\Sigma}_{k}|}}\,.  \label{eq:e2}
		\end{equation}
$|\VEC{\Sigma}_{k}|$ is the determinant of $\VEC{\Sigma}_{k}$, the
$\VEC{\mu}_k$ and $\VEC{\Sigma}_k$ are the mean vector and
co-variance of the $k$-th Gaussian respectively.

The parameters of GMM, i.e. $P_{k}$, $\VEC{\mu}_{k}$ and $\VEC{\Sigma}_{k}$, 
can be determined
from the data by an unsupervised machine learning technique, namely the
Expectation-Maximization (EM) algorithm \citep{NR3}, which assumes no prior 
knowledge
about 
the clustering structures.
For $N$ data points $\VEC{x_{i}}$, where $i=1\ldots N$, the EM 
algorithm starts from an initial guess and learns GMM parameters from the data.  
The steps involved are:
\begin{itemize}
	\item Guess starting values for $\VEC{\mu}_{k},
		\VEC{\Sigma}_{k}$, and $P_{k}$
	\item Expectation-step (E-step): calculate
		$P(\VEC{x}|\VEC{\mu}_{k}, \VEC{\Sigma}_{k})$ and $P(\VEC{x})$
using
		Equations (\ref{eq:e1})~and~(\ref{eq:e2}).
	\item Maximization-step (M-step): Update model parameters $
		\VEC{\mu}_{k}, \VEC{\Sigma}_{k}$ using \begin{eqnarray}
\VEC{\mu_{k,{\rm
			new}}}&=&\frac{\sum_{i=1}^{N} \VEC{x}_{i} P_{k}
P(\VEC{x}_{i}|
\VEC{\mu}_{k}, \VEC{\Sigma}_{k})/P(\VEC{x}_{i})}{ \sum_{i=1}^{N} P_{k}
P(\VEC{x}_{i}|
			\VEC{\mu}_{k}, \VEC{\Sigma}_{k}) /P(\VEC{x}_{i})} \,,\\
\VEC{\Sigma_{k,{\rm
			new}}}&=&\frac{\sum_{i=1}^{N}
(\VEC{x}_{i}-\VEC{\mu}_{k}) \otimes
			(\VEC{x}_{i}-\VEC{\mu}_{k}) P_{k} P(\VEC{x}_{i}|
\VEC{\mu}_{k},
			\VEC{\Sigma}_{k}) /P(\VEC{x}_{i})}{ \sum_{i=1}^{N} P_{k} P(\VEC{x}_{i}|
\VEC{\mu}_{k},
			\VEC{\Sigma}_{k}) /P(\VEC{x}_{i})}\,, \\ P_{k,{\rm
			new}}&=&\frac{1}{N}\sum_{i=1}^{N} P_{k} P(\VEC{x}_{i}|
\VEC{\mu}_{k} ,
			\VEC{\Sigma}_{k}) /P(\VEC{x}_{i})\,, \end{eqnarray}
where $i$ is the index for data points. $\otimes$ is the symbol for `outer
product', i.e. for any vectors $\VEC{x}$ and $\VEC{y}$, $\VEC{x}\otimes\VEC{y}$ is
the matrix, of which the $\nu$-th row $\mu$-th column element is the product of
the $\nu$-th element of $\VEC{x}$ and the $\mu$-th element of $\VEC{y}$.

		\item Repeat the EM steps, until the
			total likelihood $\Lambda$ converges, where
\begin{equation}
				\label{eq:likdef}
				\Lambda=\prod_{i=1}^{N} P(\VEC{x}_{i})\,. 
\end{equation}
\end{itemize}
It can be shown that the above iteration of the EM algorithm
increases the total likelihood $\Lambda$, and the iterations always converge 
\footnote{In real situations, due to the finite numerical precision, this is
not always true.}.

 \begin{figure} \centering \includegraphics[totalheight=2in]{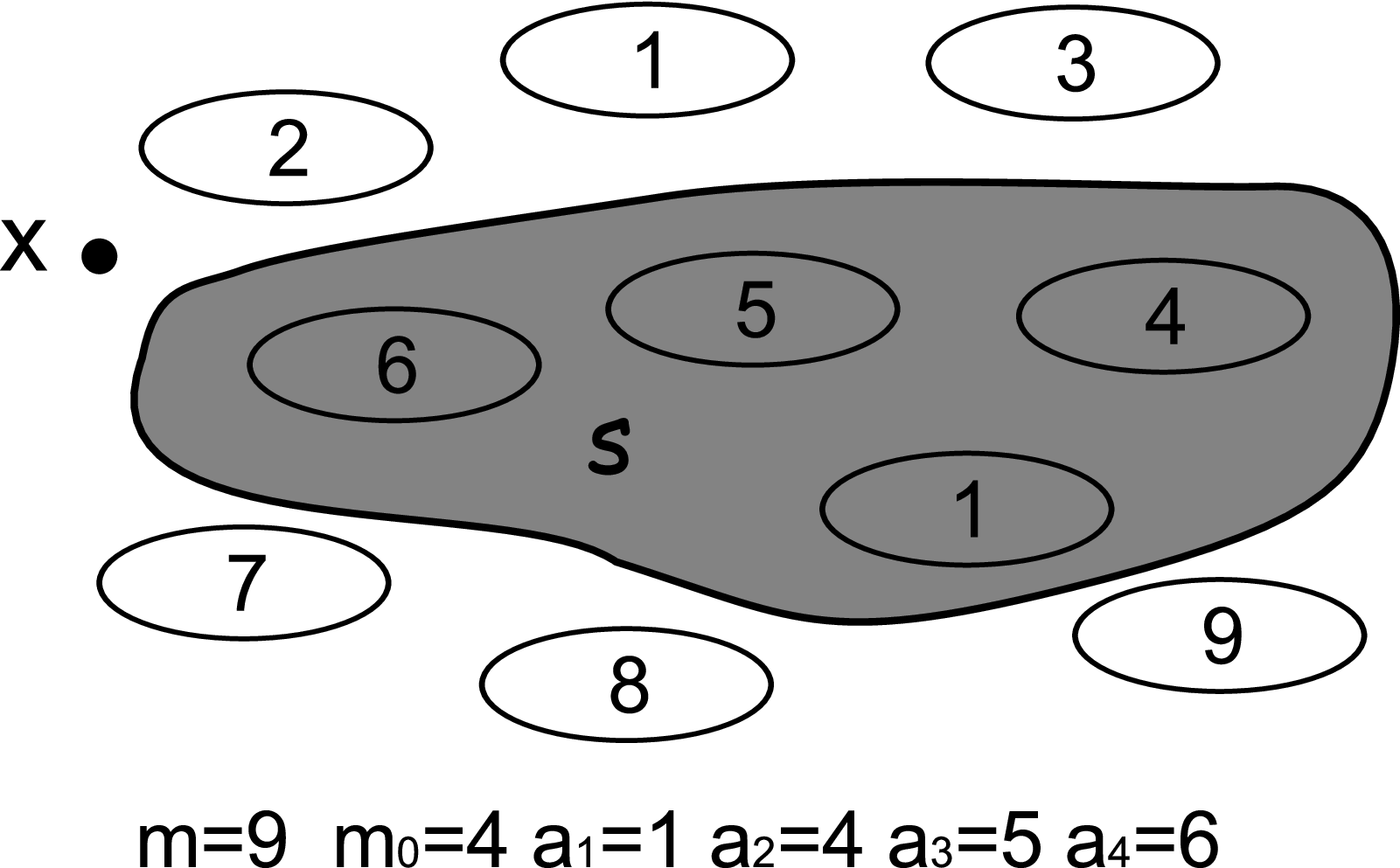}
 \caption{An
	 illustration of the definition of the GMM and data
	 clustering. The $\VEC{x}$ indicates a data point. The
	 index for each Gaussian is labeled.  Here the total number of
	 Gaussian clusters is $m=9$, where four of them belong to the
	 subset $\cal S$ ($m_0=4$), as indicated by the shaded area,
	 thus $\VEC{x} \notin {\cal S}$ \label{fig:clsill}}  \end{figure}

With the GMM and its parameters, one can infer the association of any
data point $\VEC{x}$ with these clusters. One can also ask if $\VEC{x}$ belongs
to a certain subset $\cal S$, which contains $m_{0}$ Gaussian clusters out
of a total of $m$ Gaussian clusters. An illustration of the definitions
is presented in \FIG{fig:clsill}, where the indexes of Gaussian
clusters in $\cal S$ are $a_{\rm k}$, and $k=1\dots m_{0}$.
The question of association can be answered via the standard likelihood
ratio test.	The Neyman-Pearson lemma \citep{Kassam88} claims that
the most powerful test for the binary hypotheses: \begin{equation}
	\left\{\begin{array}{c}
		\textrm{$H_1$: $\VEC{x}$ belongs to subset $\cal S$\,;}
		\\ \textrm{$H_0$: $\VEC{x}$ belongs to other clusters \,,}
	\end{array}\right.  \label{eq:stahy}
\end{equation} is to compare the logarithmic likelihood ratio
$\log R_{\cal S}$ against a statistical decision threshold $\eta$,
i.e. choose $H_{0}$, if $\log R_{\cal S} \ge \eta$, otherwise choose
$H_{1}$. According to the GMM, the logarithmic likelihood ratio $R_{\cal
S}$ is \begin{equation}
	\log R_{\cal S}= \log\left(\frac{	\sum_{k\in {\cal
	S}}P_{k}P(\VEC{x}| \VEC{\mu}_{k}, \VEC{\Sigma}_{k})
	}{     \sum_{k\notin {\cal S}} P_{k'} P(\VEC{x}|
	\VEC{\mu}_{k'}, \VEC{\Sigma}_{k'}) } \frac{	 \sum_{k'
	\notin {\cal S}}P_{k'}}{\sum_{k\in {\cal S}} P_{k} }
	\right)\,,\label{eq:likelyrank}
\end{equation} where summation $\sum_{k \in {\cal S}}$ sums over the
index $k$ for those clusters in the subset $\cal S$ and  $\sum_{k \notin
{\cal S}}$ sums over the complementary set of $\cal S$, i.e. those clusters
not in the subset $\cal S$.

The number of Gaussian clusters $m$, as an input parameter, can be
determined using statistical methods. In practice, one usually starts
from $m=1$, and then increases $m$. As $m$ is increased, one can
describe the data better.  To avoid over-fitting, it is necessary to
check the modelling using the multi-dimensional Kolmogorov-Smirnov test
(K-S test). Similar to a 1-D K-S test, the multi-dimensional K-S test is
used to test whether two data set differ significantly from each other
or whether a data set differs from a known distribution. The statistic
($\cal D$) for K-S test is the maximal difference between the cumulative
distribution of two data sets or between the data and the model. However,
the cumulative distribution of multi-dimensional data is not well defined,
thus it was proposed to compute such `cumulative distribution' for any
possible order and then calculate the $\cal D$. For example, in order to
determine the maximal difference $\cal D$ between data and data or between
data and distribution, it is necessary to check the cumulative probability
for all of the four cases $(x<x_i, y<y_i), (x<x_i, y>y_i), (x>x_i,
y<y_i)$, and $ (x>x_i, y>y_i)$ for any data point $(x_i, y_i)$ belonging
to the two dimensional data set $\VEC{x}=(x_i,y_i), i\in [1,N]$. For
a multi-dimensional K-S test, the statistical threshold and p-values are
usually calculated numerically via Monte-Carlo simulation, which
generates the mock data sets and calculates the null-hypothesis distribution
of $\cal D$ accordingly. We refer readers to \cite{PE83} and \cite{FF87}
for more details of such tests.

In summary, the technique of using the GMM to describe a data distribution
and to determine the data association is given in the following recipe:

1. Determine the parameter space and form the data vector $\VEC{x}$ for the data
set.

2. Guess the number of clusters $m$ and their initial parameters, i.e.  
$\VEC{\mu}_{k}$ and $\VEC{\sigma}_{k}$, where $k=1\dots m$.

3. Use the EM algorithm to find the true model parameters. 

4. Check if GMM describes the data distribution well enough using
the multi-dimensional Kolmogorov-Smirnov test. Increase the number of Gaussian 
clusters, if
the test fails.

5. Once the model parameters are found, use \EQ{eq:likelyrank} to determine the
data association. 

We
present two examples in next section to show its applications.

\section{Application} \label{sec:app}
\subsection{Application to pulsar classification}
\label{sec:psrapp}

As the first example, we apply GMM to the well-known pulsar \PPD
diagram to find a quantitative description for the pulsar distribution.
We also seek the `empirical MSP definition' here, especially because a precise
definition for MSPs using their periods and period derivatives is not available
yet.

The standard picture for pulsar evolution contains several major stages,
i) the birth of pulsar in a supernova explosion, ii) the spin-down
of pulsar due to radiation energy loss, iii) the pulsar death due to
the decrease in radiation power, and possibly for some binary system,
iv) the pulsar recycling by accretion induced spin-up.  After birth in
the supernova, the \emph{young pulsars} usually have short periods and
large period derivatives. They occupy the upper left part of the \PPD
diagram, and are frequently associated with supernova remnants. As the
pulsars age, they slow down, while, for those pulsars with breaking
index $n=2-\ddot{P}P/\dot{P}^2>2$, their $\dot{P}$ also decreases.
The pulsars then enter the main population in the centre of the diagram,
referred to as the \emph{normal pulsars}. Eventually such spin-down
causes the \emph{death} of pulsars, i.e.\, the radiation of pulsar ceases
or becomes too weak to detect. Pulsars may also get \emph{recycled}
via the accretion process \citep{BV91}. The \emph{millisecond pulsars}
(MSPs), which occupy the lower left corner of the \PPD diagram, are
commonly believed to form due to such a spin-up of a normal pulsar via
accretion materials from a companion stars. A continuum of pulsars from the MSP
population to the normal pulsar population is already observed, where
the intermediate population are referred to as \emph{mildly recycled
pulsars}. There are also pulsars occupying the upper right part of the
\PPD diagram. The origin and evolution of these \emph{high magnetic
field pulsars} is still unclear.

\begin{figure*} \centering \includegraphics[totalheight=5.5in]{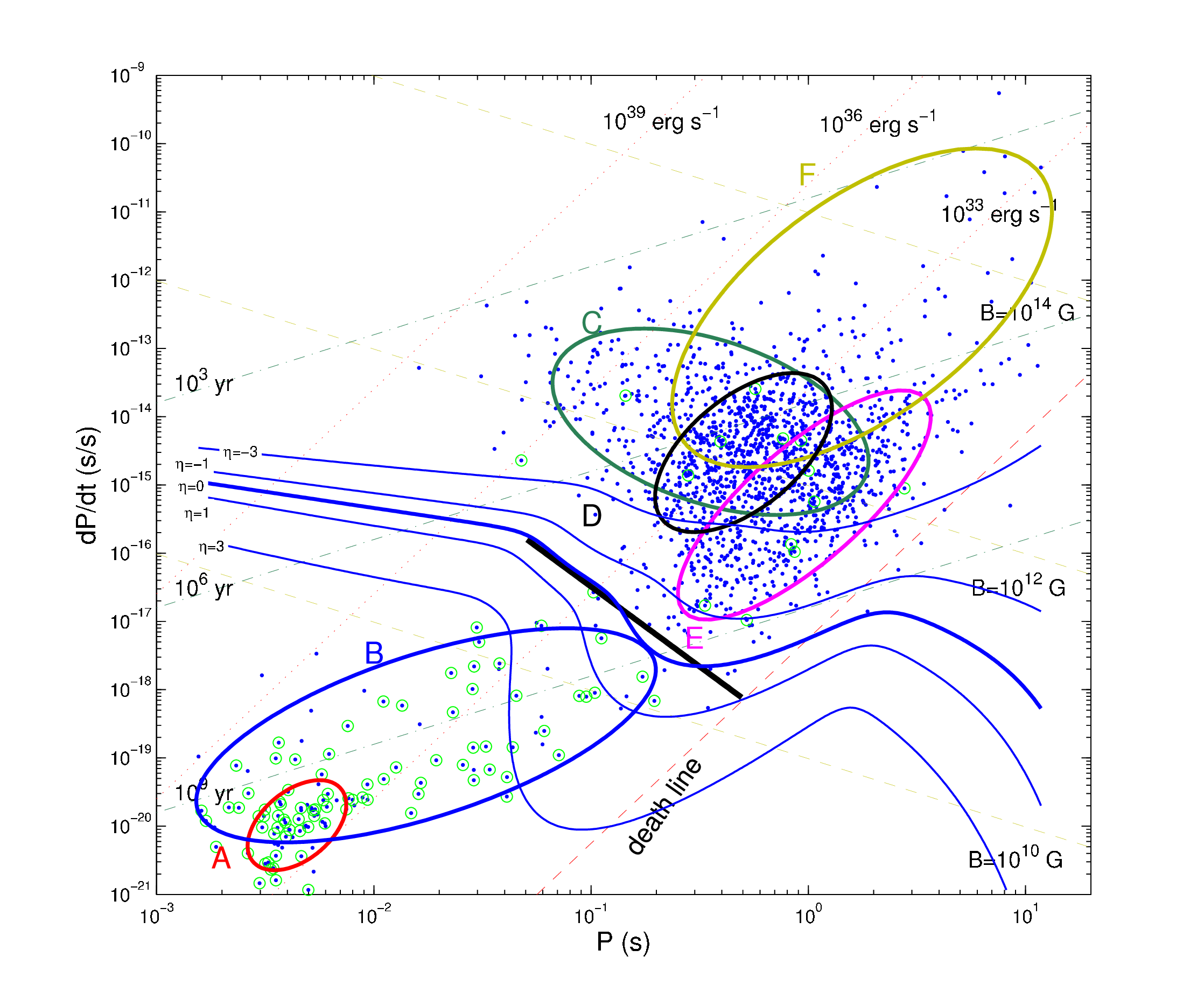} \caption{ The
	pulsar $P$-$\dot{P}$ diagram.	Elliptic curves are $2-\sigma$ contours for the
	six Gaussian clusters found by GMM. Dots are all detected pulsars from the
	ATNF catalogue, and those with circles are binaries. Blue curves with
	numerical labels on the left ends are contours for equal logarithmic
	likelihood $\log(R_{\cal S})$. The definition of MSP according to our model is
	$\log R_{\cal S}\leq0$, which is plotted with thick blue line. The linear
	approximation of $\log R_{\cal S}=0$ with $50\,\textrm{ms}\le P
	\le0.5\,\textrm{s}$ (i.e. \EQ{eq:defmsp}) is plotted using a thick black 
	straight line.  Lines with equal magnetic field
	strength, spin-down power, characteristic age as well as the death line are
	also plotted.  Data are from the ATNF pulsar catalogue (\citealt{MHT05},
	http://www.atnf.csiro.au/research/pulsar/psrcat/).  \label{fig:psrdis}}
\end{figure*}

We use pulsar $P$ and $\dot{P}$ values from the ATNF pulsar catalogue
\citep{MHT05}.  The
distribution of pulsars is plotted in \FIG{fig:psrdis}. The distribution of
the whole pulsar population does not follow a Gaussian distribution, but we
can approximate the overall distribution with multiple Gaussian clusters, i.e.
the GMM is still valid for the modelling purposes. We apply the GMM in the \PPD
diagram, so our parameter vector $\VEC{x}$ is \begin{equation}
	\VEC{x}=\log\left(\begin{array}{c} P \\ \dot{P} \end{array}\right)\,.
	\end{equation}

We directly apply the GMM to the data set. The EM algorithm for GMM is
stable so the EM algorithm converges for most initial values, although
the EM algorithm does not guarantee that it converges to the best global
maximum of the total likelihood $\Lambda$. In order to attain the global
maximum, initial GMM parameters are generated randomly from a uniform
distribution covering the whole \PPD diagram (in particular,we choose
the range of $P$ from $10^{-4}$ to 20 s, and the range of $\dot{P}$
from $10^{-22}$ to $10^{-10}$).  We then use these initial parameters
in the EM algorithm.  We repeat this process $10^4$ times to determine
the best global model parameters giving the largest likelihood value. In this case, the 
probability to converge to the best model is more than 30\% for all
guesses.
The GMM is tested using the multi-dimensional Kolmogorov-Smirnov test
\citep{PE83,FF87}, for which the $p$-value is chosen to be 95\%. To pass
such test, we need six different Gaussian components in the GMM.

\begin{table}
\caption[Numerical values of the GMM parameters for the pulsar distribution in 
the $P-\dot{P}$ space.  ]{Numerical values of the GMM parameters for the pulsar 
distribution in the $P-\dot{P}$ space. The definitions are in Equations 
(\ref{eq:e1}) and (\ref{eq:e2}). The mixture weights, $P_{A}\dots P_{F}$, as 
probabilities, are dimensionless.  The central vectors of clusters,
$\VEC{\mu}_{A} \dots \VEC{\mu}_{F}$, have the same units with respect to
$\VEC{x}$, i.e.  $(\log($s$)$, $\log($s/s$))$. The covariance  matrices
$\VEC{\Sigma}_{A}\dots\VEC{\Sigma}_{F}$ are in the linear space of
$\VEC{x}\bigotimes\VEC{x}$, so their units are $\left(\begin{array}{c c}
	\log(\textrm{s})^2 & \log(\textrm{s}) \log(\textrm{s/s}) \\ \log(\textrm{s})
	\log(\textrm{s/s}) & \log(\textrm{s/s})^2 \end{array} \right).$
	\label{tab:parappd}} \begin{center} \begin{tabular}{c|l}\hline\hline Name  &
		Value \\ \hline $P_{A}$ & 0.0326 \\ $P_{B}$ & 0.0403 \\ $P_{C}$ & 0.2474 \\
		$P_{D}$ & 0.3170 \\ $P_{E}$ & 0.3337 \\ $P_{F}$ & 0.0290 \\ \hline
		$\VEC{\mu}_{A}$	& $\left(\begin{array}{l l} $-$2.3541& $-$19.9847
		\end{array} \right)$ \\ $\VEC{\mu}_{B}$	& $\left(\begin{array}{l l}
			$-$1.7597& $-$18.6687 \end{array} \right)$ \\ $\VEC{\mu}_{C}$	&
			$\left(\begin{array}{l l} $-$0.4502& $-$14.0749 \end{array} \right)$ \\
				$\VEC{\mu}_{D}$	& $\left(\begin{array}{l l} $-$0.2983& $-$14.5266
				\end{array} \right)$ \\ $\VEC{\mu}_{E}$	& $\left(\begin{array}{l l}
					$-$0.0179& $-$15.2945 \end{array} \right)$ \\ $\VEC{\mu}_{F}$	&
					$\left(\begin{array}{l l} 0.2477& $-$12.4064 \end{array} \right)$ \\
						\hline $\VEC{\Sigma}_{A}$ & $\left(\begin{array}{l l} 0.0172& 0.0229
							\\0.0229& 0.1451 \end{array}\right)$ \\ $\VEC{\Sigma}_{B}$ &
							$\left(\begin{array}{l l} 0.3732& 0.3388 \\0.3388& 0.8204
							\end{array}\right)$ \\ $\VEC{\Sigma}_{C}$ & $\left(\begin{array}{l
								l} 0.1764& $-$0.1382 \\$-$0.1382& 0.6213 \end{array}\right)$ \\
								$\VEC{\Sigma}_{D}$ & $\left(\begin{array}{l l} 0.0552& 0.0886
									\\0.0886& 0.4530  \end{array}\right)$ \\ $\VEC{\Sigma}_{E}$ &
									$\left(\begin{array}{l l} 0.1131& 0.2524 \\0.2524& 0.9418
									\end{array}\right)$ \\ $\VEC{\Sigma}_{F}$ &
									$\left(\begin{array}{l l} 0.2547& 0.4106 \\0.4106& 1.8187
									\end{array}\right)$ \\ \hline \hline \end{tabular}
								\end{center} \end{table}

We check the `robustness' of GMM parameters using an algorithm similar
to the bootstrap method. The original \PPD data are re-sampled with
replacements to form $10^2$ simulated data sets, i.e.\, any data point
has the same probability of being sampled at any time. The EM algorithm is
then applied to these newly simulated data sets and check the stability of
GMM parameters as a function of the total number of data points in each
simulated data set. We find that there is no significant change in the
structure of the Gaussian clusters, if we randomly remove less than 3\% of the data.  
We
also have checked the GMM parameter stability with respect to the Shklovskii
effect \citep{Shk70}, an effect that increases the observed period derivative
of the pulsar due to its transverse velocity. The differences between
GMM parameters for pulsar distributions with and without correcting the
Shklovskii effect are less than 1\% (the corrected $\dot{P}$ values
are also from the ATNF catalogue). Since the Shklovskii correction is
not substantial for the GMM, we ignore it in the rest of the discussions.  One 
needs to
re-do the above analysis to further check the stability and robustness
of the model parameters, when more pulsar data becomes available.


As plotted in \FIG{fig:psrdis}, in order to describe the pulsar
distribution in the \PPD diagram, six Gaussian clusters are required, the
parameters of which are listed in \TAB{tab:parappd}. It is clear that two
Gaussian clusters (components `A' and `B') are required for describing
the MSP distribution and four clusters (components `C', `D',`E', and `F')
are required for normal pulsars.  We calculate the likelihood ratio
according to \EQ{eq:likelyrank}, where the set $\cal S$ contains the two
MSP clusters. We plot the equal likelihood ratio contours corresponding
to $R_{\cal S}=\eta$ in \FIG{fig:psrdis}.  Any of these curves divides
the \PPD diagram into two regions, one for MSPs and the other for normal
pulsars. We can now empirically define MSPs as pulsars satisfying $R_{\cal
S} \geq0$. We can derive a linear approximation in $\log P$ to $R_{\cal
S}=0$. In order to avoid having regions, where only very few pulsars are
present, the linear approximation can be confined to the interesting
range of $P$, i.e. the linear approximation is calculated for $P$ from
$50$ ms to $0.5$ s as \begin{equation} \frac{\dot{P}}{10^{-17}}
	\leq3.23\left( \frac{P}{100\, \textrm{ ms}}\right)^{-2.34}\,.
	\label{eq:defmsp} \end{equation} This can be seen as an empirical
	`definition' for MSPs.

Individual Gaussian clusters may be artifacts, due to an intrinsic pulsar
distribution that is non-Gaussian. In this case, multiple components are
required to describe the distribution. Such non-Gaussian distribution may
come from selection effects in pulsar searching or pulsar evolutionary
mechanisms. However, we would like to mention a few interesting features
here, which may agree with other evidence.  

There are two possible MSP clusters. As shown in \FIG{fig:psrdis}, the
principal axes with maximal eigenvalue for component `A' is parallel
to constant energy losing rate lines (defined by $\dot{E}=4\pi^2
I\dot{P}P^{-3}=3.9\times 10^{46} \left({P}/{\textrm{ s}}\right)^{-3}
{\dot{P}}$\,{erg}) , while the principal axes with maximal eigenvalue of
component `B' is parallel to the equal characteristic age lines (defined
by $\tau_{\rm c}=0.5\, P \dot{P}^{-1}$).  Such different directions of
eigenvectors may indicate that the MSPs have two different origins or
evolutionary tracks, which is supported by the fact that that nearly
all component-`A' pulsars are He-white dwarf binaries, while the major
population in component `B' are CO-white dwarf binaries \citep{Tauris11}.

Component `B' contains mildly recycled pulsars, which have
larger period and higher magnetic field than other MSPs. According to
the GMM, there is no statistical evidence for a separate cluster of
mildly recycled pulsars in the \PPD diagram. This may indicate a `smooth
spectrum' for the amount of accreted mass for fully recycled and mildly
recycled MSPs, otherwise we would expect a more complex structure in
component `B'.

There are possibly four normal pulsar clusters. One each for high magnetic field
pulsars (component `F'), old pulsars close to the death line (component
`E'), young pulsars (component `C'), and middle age pulsars (component
`D').  Principal axes directions of these ellipses for middle age pulsars,
old pulsars, and high magnetic field pulsar are nearly parallel (within
$\sim 3^\circ$). Such agreement indicates certain common mechanisms
among these pulsars, because the probability of three random Gaussian
clusters having parallel principal axes within $3^\circ$ is by chance
small as $(3/90)^2 \simeq 10^{-3}$.

The young pulsar ellipse `C' is clockwise rotated by $\sim 15^\circ$
with respect to the average principal axis directions of ellipses
`D', `E', and `F' \footnote{Because of the unequal X-Y scales, the
$15^\circ$ rotation is distorted in \FIG{fig:psrdis} visually.}.
Such a rotation may come from the selection effect that it is easier
to find bright pulsars. Bearing this selection effect in mind, this
rotation may indicate that the old pulsars and the young pulsars have
different radiation or evolution mechanisms, which is supported by both the 
timing and polarization properties. Timing results \citep{ELS11} show that 
pulsars with glitch behavior correlates with clusters `C', while the 
polarization measurements \citep{WJ08} indicate that the polarization behaviors 
for high and low $\dot{E}$ pulsar are different.

We also notice that the ellipse for high magnetic field pulsars is quite
extended instead of being localized to only high magnetic field pulsars.
It covers regions of both young pulsars and middle age pulsars. This
indicates potential links between the high magnetic field population
and normal pulsars, for which, some observational evidence already
suggests that high magnetic field pulsars may evolve from young pulsars
\citep{Lyne04, LZ04, ELKM11}.

\subsection{Application to classification of {\it Fermi} point sources}
\label{sec:ferapp}

\begin{figure*} \centering \includegraphics[totalheight=3in]{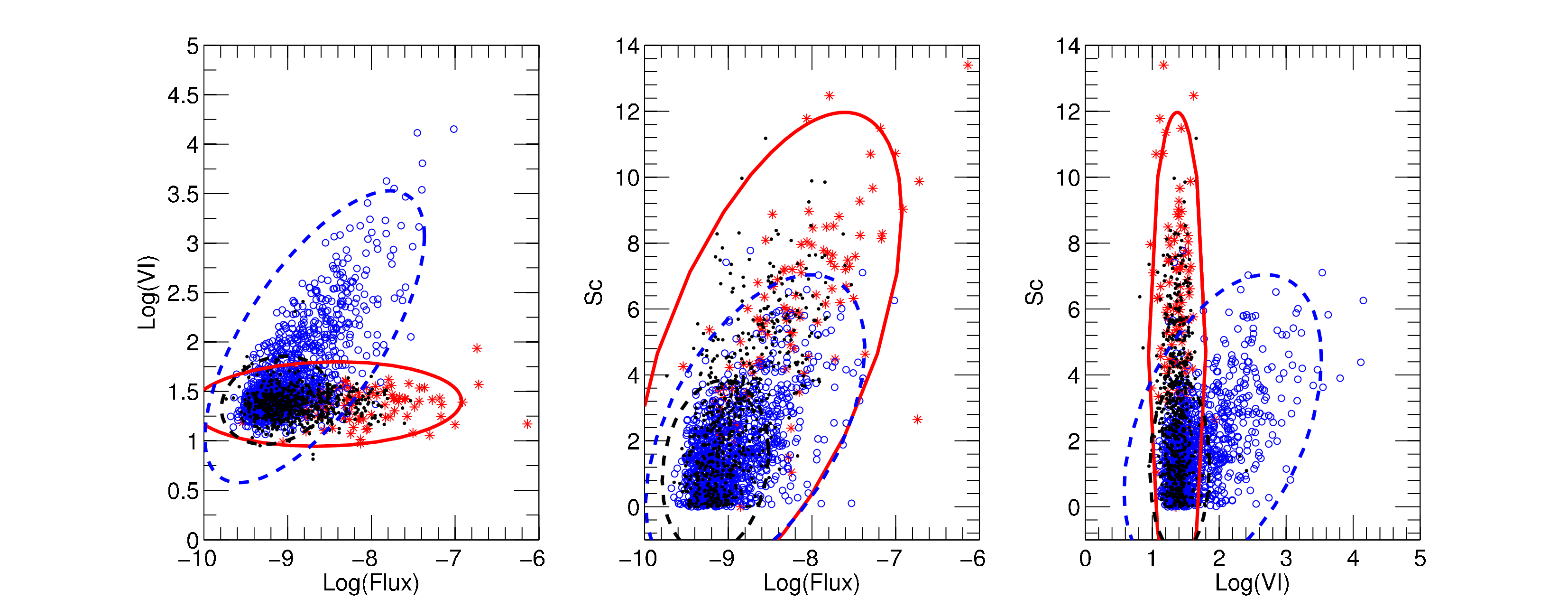}
\caption{The projected Gaussian clusters and source distribution. Blue
`$\circ$' symbols are AGs, red `$*$' symbols are pulsars, un-associated
sources are plotted with black solid dots. The units we use for
each axis is identical to that in the 2FGL catalog, where the \FTH\
takes the unit of $\rm cm^{-2}\, s^{-1}$. The significance of
curved spectrum and variability index are dimensionless according
to the 2FGL catalog. There is a 3-D movie of this figure available at
http://www.mpifr-bonn.mpg.de/staff/kjlee/gmm.html.  \label{fig:distr}
} \end{figure*}

We now apply the GMM to the 2FGL catalog to demonstrate the
use of GMM to rank candidates according to likelihood.  As already pointed out
by \cite{LAT1FG} and \cite{Fermi11}, one can use the variability and the
spectrum information to classify \textit{Fermi} point sources.  In the 2FGL catalog,
the Variability\_Index ($V\!I$) and Signif\_Curve (significance of the fit
improvement using a curved spectrum, $Sc$) are used to describe the variability
and spectral shape. Figures \ref{fig:distr} show that
\textit{Fermi} point sources form two distinctive classes, the pulsars and the
active galaxies (AGs), where pulsars usually have smaller $V\!I$ but larger $Sc$
compared to AGs.

Although it is straightforward in classifying, one still needs to
be careful using parameters $V\!I$ and $Sc$, since both $V\!I$
and $Sc$ correlate with the Test Statistics (see \citealt{LAT1FG}
for the definition). Such correlation smears clustering structures
at low detection significance. One way to deal with this situation
is by re-defining $V\!I$ and $Sc$ to reduce the correlation
\citep{Fermi11}. Here, we take an alternative approach, in which the
gamma-ray flux is included as an additional parameter to directly correct
the correlation. In other words, instead of using only $V\!I$ and $Sc$,
we use three variables, $V\!I$, $Sc$, and integral gamma-ray flux from
1 to 100 GeV, `Flux1000' (\FTH) to classify data. Figures \ref{fig:distr} show 
that pulsars and AGs have different dependence
in gamma-ray flux, which helps us in data classification.

We now turn to details of our setup for the GMM. The parameter space we used in
the GMM algorithm is spanned by $Sc$, base-10 logarithms of $V\!I$ and
\FTH, i.e.  the components of data vector $\VEC{x}$ are: \begin{equation}
	\VEC{x}=\left(\begin{array}{c} \log F_{1000} \\ \log V\!I \\ Sc
	\end{array}\right)\,.  \label{eq:xdef} \end{equation}

The distribution of $\VEC{x}$ for the all 2FGL catalog sources is plotted in
\FIG{fig:distr}. The GMM is
then applied to the whole population to determine the model parameters.
To test the robustness of the algorithm and to check if the result
is sensitive to initial values, we run the EM algorithm with randomly
generated initial values as in the previous example. We get three
Gaussian clusters, where one of the Gaussian clusters overlaps with the AGs, one
overlaps with pulsars, and one is for sources with low fluxes. The GMM
parameters turn out to be insensitive to initial values. We note that extra
Gaussian components are needed to model the details of the low flux branch to
pass the multidimensional Kolmogorov-Smirnov test.  Such a requirement is simply
due to the fact that $Sc\ge0$, where such a boundary needs more Gaussian
components to approximate the distribution. We have also checked that including
these extra components will not significantly alter our results of source
classification, thus for simplicity we prefer to use only three Gaussian
clusters here.  The projected 3-$\sigma$ contours of clusters are shown in
\FIG{fig:distr}.  The GMM parameters, i.e.  $\VEC{\mu}$ and $\VEC{\Sigma}$, are
given in \TAB{tab:para}. 

%

\begin{table}

\caption[GMM parameters for point source distribution in the
	2FGL catalog]{GMM parameters for point source distribution in the
	2FGL catalog. Similar to \TAB{tab:parappd}, the $P_{1}\dots P_{3}$ are dimensionless.  
	The
	vectors
$\VEC{\mu}_{1} \dots \VEC{\mu}_{3}$ have the same units with respected
to $\VEC{x}$, i.e. $\left(\log(\textrm{cm}^{-2}\textrm{s}^{-1}),
\textrm{dimensionless}, \textrm{dimensionless}\right)$. The matrices 
$\VEC{\Sigma}_{1}\dots\VEC{\Sigma}_{3}$ take units of $\left(\begin{array}{c c
c}
\log(\textrm{cm}^{-2}\textrm{s}^{-1})^2 & \log(\textrm{cm}^{-2}\textrm{s}^{-1})
& \log(\textrm{cm}^{-2}\textrm{s}^{-1}) \\
\log(\textrm{cm}^{-2}\textrm{s}^{-1}) & \textrm{dimensionless} &
\textrm{dimensionless}\\
\log(\textrm{cm}^{-2}\textrm{s}^{-1}) & \textrm{dimensionless} &
\textrm{dimensionless}
\end{array}\right)$ }

\label{tab:para}
\begin{center}
\begin{tabular}{c|ccc}\hline\hline
	Name  & Value   &  \\
	\hline
	$P_{1}$ & 0.2856 \\
	$P_{2}$ & 0.2290\\
	$P_{3}$ & 0.4855 \\
	\hline
	$\VEC{\mu}_{1}$	& $\left(\begin{array}{l l l}$-$8.5330 & 1.3732 &
4.6445\end{array}\right)$	\\
	$\VEC{\mu}_{2}$	& $\left(\begin{array}{l l l} $-$8.6773 & 2.0533	& 2.4719
\end{array}\right)$	\\
	$\VEC{\mu}_{3}$	& $\left(\begin{array}{l l l} $-$9.1534 & 1.4110	& 1.2071
	\end{array}\right)$	\\
	\hline
	$\VEC{\Sigma}_{1}$  & $\left(\begin{array}{l l l} 2.8686\textrm{e}$-$01	
		&5.9415\textrm{e}$-$03
&7.5103\textrm{e}$-$01\\
		5.9415\textrm{e}$-$03	&2.0322\textrm{e}$-$02
&1.9244\textrm{e}$-$03\\
		7.5103\textrm{e}$-$01	&1.9244\textrm{e}$-$03
&5.9513\textrm{e}$+$00\\
	\end{array} \right)$ \\
$\VEC{\Sigma}_{2}$  & $\left(\begin{array}{l l l} 1.9069\textrm{e}$-$01	
	&1.4785\textrm{e}$-$01	&3.3875\textrm{e}$-$01\\
	1.4785\textrm{e}$-$01	&2.4183\textrm{e}$-$01	&3.2970\textrm{e}$-$01\\
	3.3875\textrm{e}$-$01	&3.2970\textrm{e}$-$01	&2.3165\textrm{e}$+$00\\
	\end{array} \right)$ \\
$\VEC{\Sigma}_{3}$  & $\left(\begin{array}{l l l} 4.5017\textrm{e}$-$02	
	&5.9696\textrm{e}$-$03	&2.9135\textrm{e}$-$02\\
	5.9696\textrm{e}$-$03	&2.2089\textrm{e}$-$02	&3.3940\textrm{e}$-$03\\
	2.9135\textrm{e}$-$02	&3.3940\textrm{e}$-$03	&7.7004\textrm{e}$-$01\\
	\end{array} \right)$ \\
		\hline \hline
\end{tabular}
\end{center}
\end{table}

\begin{figure} \centering \includegraphics[totalheight=3.3in]{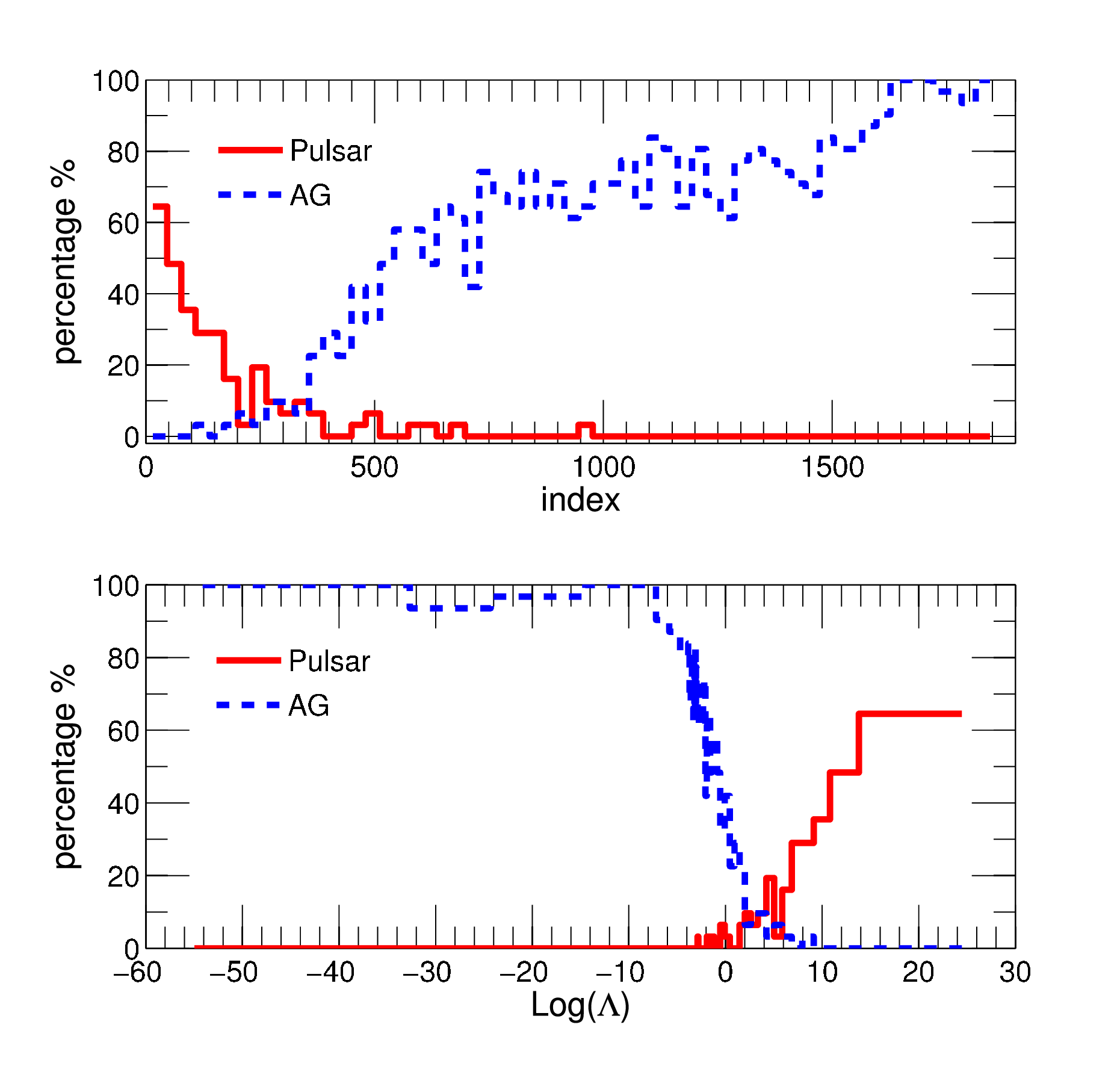}
	\caption{\textit{Upper panel:} The proportion of pulsars and AGs
	as functions of ranking index. We bin all the 2FGL catalog sources into
	60 bins according to the ranking index. The X-axis is the average
	index value, and the Y-axis is the   proportion of pulsars/AGs
	in the bin. The solid line is for pulsars, while the
	dashed line is for AGs.  One can see that our algorithm
statistically ranks pulsars higher than AGs.	\textit{Lower panel:} The
proportion of pulsars and AGs as functions of logarithms of pulsar
likelihood. The X-axis is the average logarithm of pulsar likelihood,
and the Y-axis is the proportion of pulsars/AGs in each bin. Due to
our definition for the likelihood, the proportion of pulsars and AGs
equals to each other are at $\log R_{\cal S}=0$, as expected.
\label{fig:prob}} \end{figure}

Our pulsar likelihood is calculated using \EQ{eq:likelyrank} with $\VEC{\mu}_{k}$ and
$\VEC{\Sigma}_{k}$ from \TAB{tab:para}. We identify the pulsar subset $\cal S$
as the pulsar cluster, the complementary set are the AG and low flux
cluster. The reason we exclude the low flux cluster from subset $\cal S$ is to
suppress candidates with low detection significances.

We sort the 2FGL catalog point sources according to the pulsar
likelihood. The whole sorted source list can be found in the online supplement 
materials of this paper. The first 5
and the last 5 sources are given as examples in \TAB{tab:samp}.

In order to verify the validity of our ranking technique,
we checked the ranking index of known pulsars and other
known sources in our list.  To identify pulsars, we used the `Public List
of LAT-Detected Gamma-Ray Pulsars'\footnote{The list is available at
https://confluence.slac.stanford.edu/display/\-GLAMCOG/Public+List+of+LAT-Detected+Gamma\-Ray+Pulsars},
which lists the pulsars that have been detected as pulsed gamma-ray
sources with the Fermi LAT up to now. For AGs, the association
information is from the 2FGL catalog, i.e. the `CLASS1' entry.
Such a comparison is legitimate, since we do not use any information
about known pulsar and AG population in the process of finding the
model parameters. The comparisons between the ranking results and
known population are shown in \FIG{fig:prob}. We can see that the
ranking technique separates the pulsar population and AG population.
The proportion of pulsars decreases and the proportion of the AGs
increases with the ranking index respectively.  The top 5\% sources
contain 50\% known pulsars, the top 50\% contain 99\% known pulsars, and
no known active galaxy appears in the top 6\%.  As predictions, we list
the potential pulsar candidates in \TAB{tab:cand}, which according to
our likelihood are highly pulsar-like objects with no associated sources
in the 2FGL catalog. We hope future pulsar searching will benefit from
this information.

\begin{table*} \caption{A sample of the source list sorted according to our
	likelihood. In this list, 2FGL name and corresponding 1FGL name of the sources
	are given. The source associations are from the 2FGL, and $\log(R_{\cal S})$
	is our logarithmic likelihood. \label{tab:samp} }
	\begin{center} \begin{tabular}{ccccccc}\hline\hline Index& 2FGL name &1FGL
		name&  RA (J2000) & DEC (J2000) & Association & $\log R_{\cal S}$\\ \hline
		1 & 2FGL J0633.9$+$1746 & 1FGL J0633.9$+$1746 & 06:33:55 & $+$17:46:26 & PSR J0633$+$1746 & $55.8$\\
		2 & 2FGL J0835.3$-$4510 & 1FGL J0835.3$-$4510 & 08:35:21 & $-$45:10:45 & PSR J0835$-$4510 & $52.6$ \\
		3 & 2FGL J1801.3$-$2326e & & 18:01:22 & $-$23:26:24 & SNR G006.4$-$00.1 & $38.2$ \\
		4 & 2FGL J1836.2$+$5926 &  & 18:36:16 & $+$59:26:01 & LAT PSR J1836$+$5925 & $33.0$ \\
		5 & 2FGL J0007.0$+$7303 & 1FGL J0007.0$+$7303 & 00:07:06 & $+$73:03:16 & LAT PSR J0007$+$7303 & $30.0$ \\
				\ldots & \ldots & \ldots & \ldots & \ldots & \ldots & \ldots \\
				1869 & 2FGL J0238.7$+$1637 & 1FGL J0238.6$+$1637 & 02:38:42 & $+$16:37:26 & AO 0235$+$164 & $-112$ \\
				1870 & 2FGL J1229.1$+$0202 & 1FGL J1229.1$+$0203 & 12:29:06 & $+$02:02:30 & 3C 273 & $-119$ \\
				1871 & 2FGL J1512.8$-$0906 & 1FGL J1512.8$-$0906 & 15:12:50 & $-$09:06:12 & PKS 1510$-$089 & $-139$ \\
				1872 & 2FGL J1224.9$+$2122 & 1FGL J1224.7$+$2121 & 12:24:54 & $+$21:22:48 & 4C $+$21.35 & $-176$ \\
				1873 & 2FGL J2253.9$+$1609 & 1FGL J2253.9$+$1608 & 22:53:59 & $+$16:09:09 & 3C 454.3 & $-181$ \\
		\hline \hline \end{tabular} \end{center} \end{table*}

\begin{table*}
\caption{A sample of pulsar candidates sorted according to the pulsar 
likelihood. The 2FGL name with $\dag$ indicate that there is a known pulsar 
within the error ellipses of the pointing according to the ATNF pulsar 
catalogue. The SemiMajor and SemiMinor axes are the source
positions at 68\% confidence in units of degrees. The column `Class' denotes the 
possible association of the source, where `SNR' denotes supernova remnant, `GLC' 
denotes globular cluster, and `SPP' denotes special cases which may associated 
with SNR or pulsar wind nebula \citep{LAT2FGL}.  \label{tab:cand}}
\begin{center}
\begin{tabular}{ccccccccc}\hline\hline
	Index&	2FGL name &  1FGL name & RA & DEC &SemiMajor
	& SemiMinor & $\log R_{\cal S}$ & Class\\
	& & & J2000 & J2000&$10^{-2}\,$deg & $10^{-2}\,$deg & \\ \hline
1 & 2FGL J1801.3$-$2326e  &     &  18:01:22  &  $-$23:26:24  & ---& --- & 38.2 & SNR \\ 
2 & 2FGL J1745.6$-$2858  &     &  17:45:42  &  $-$28:58:42  & 1.0& 1.0 & 29.0 & SPP \\ 
3 & 2FGL J1855.9$+$0121e  &     &  18:55:58  &  $+$01:21:18  & ---& --- & 27.5 & SNR \\ 
4 & 2FGL J0617.2$+$2234e  &  1FGL J0617.2$+$2233  &  06:17:14  &  $+$22:34:47  & ---& --- & 27.1 & SNR \\ 
5 & 2FGL J1906.5$+$0720  &  1FGL J1906.6$+$0716c  &  19:06:35  &  $+$07:20:33  & 3.5& 2.9 & 18.2 &   \\ 
6 & 2FGL J1923.2$+$1408e  &     &  19:23:16  &  $+$14:08:42  & ---& --- & 18.0 & SNR \\ 
7 & 2FGL J1045.0$-$5941  &  1FGL J1045.2$-$5942  &  10:45:00  &  $-$59:41:31  & 1.5& 1.4 & 17.7 &   \\ 
8 & 2FGL J1704.9$-$4618  &     &  17:04:59  &  $-$46:18:14  & 15.9& 11.1 & 16.6 &   \\ 
9 & 2FGL J0848.7$-$4324  &     &  08:48:45  &  $-$43:24:24  & 9.4& 6.9 & 16.5 &   \\ 
10 & 2FGL J1738.9$-$2908  &     &  17:38:57  &  $-$29:08:24  & 15.1& 6.6 & 15.7 & SPP \\ 
\\
11 & 2FGL J1819.3$-$1523  &  1FGL J1819.4$-$1518c  &  18:19:21  &  $-$15:23:29  & 7.4& 5.5 & 15.5 &   \\ 
12 & 2FGL J1747.3$-$2825c  &     &  17:47:24  &  $-$28:25:52  & 3.6& 3.2 & 15.4 &   \\ 
13 & 2FGL J1805.6$-$2136e  &     &  18:05:38  &  $-$21:36:42  & ---& --- & 15.4 & SNR \\ 
14 & \dag2FGL J1748.0$-$2447  &  1FGL J1747.9$-$2448  &  17:48:00  &  $-$24:47:04  & 2.3& 2.2 & 14.8 & GLC \\ 
15 & 2FGL J2018.0$+$3626  &     &  20:18:03  &  $+$36:26:54  & 3.7& 3.1 & 14.3 &   \\ 
16 & 2FGL J1839.0$-$0539  &     &  18:39:04  &  $-$05:39:21  & 1.7& 1.6 & 14.2 &   \\ 
17 & 2FGL J1901.1$+$0427  &     &  19:01:11  &  $+$04:27:27  & 7.1& 5.7 & 14.1 &   \\ 
18 & 2FGL J1748.6$-$2913  &  1FGL J1748.3$-$2916c  &  17:48:39  &  $-$29:13:52  & 4.1& 3.5 & 13.8 &   \\ 
19 & 2FGL J1932.1$+$1913  &  1FGL J1932.1$+$1914c  &  19:32:10  &  $+$19:13:25  & 4.2& 3.8 & 13.8 & SPP \\ 
20 & 2FGL J1847.2$-$0236  &  1FGL J1846.8$-$0233c  &  18:47:14  &  $-$02:36:40  & 7.0& 4.6 & 13.4 &   \\ 
\\
21 & 2FGL J1856.2$+$0450c  &     &  18:56:14  &  $+$04:50:16  & 8.0& 6.4 & 13.3 &   \\ 
22 & 2FGL J0858.3$-$4333  &     &  08:58:20  &  $-$43:33:34  & 9.1& 8.7 & 12.6 &   \\ 
23 & 2FGL J1521.8$-$5735  &  1FGL J1521.8$-$5734c  &  15:21:50  &  $-$57:35:53  & 3.4& 3.1 & 12.4 & SPP \\ 
24 & 2FGL J1625.2$-$0020  &  1FGL J1625.3$-$0019  &  16:25:13  &  $-$00:20:04  & 3.5& 3.2 & 12.2 &   \\ 
25 & 2FGL J0224.0$+$6204  &  1FGL J0224.0$+$6201c  &  02:24:06  &  $+$62:04:35  & 3.5& 3.1 & 12.2 &   \\ 
26 & 2FGL J1857.8$+$0355c  &  1FGL J1857.9$+$0352c  &  18:57:53  &  $+$03:55:29  & 10.0& 6.9 & 11.9 &   \\ 
27 & 2FGL J1739.6$-$2726  &     &  17:39:40  &  $-$27:26:03  & 12.3& 7.0 & 11.8 &   \\ 
28 & 2FGL J1619.0$-$4650  &     &  16:19:04  &  $-$46:50:48  & 26.4& 15.3 & 11.5 &   \\ 
29 & 2FGL J1405.5$-$6121  &  1FGL J1405.1$-$6123c  &  14:05:30  &  $-$61:21:51  & 3.5& 2.9 & 11.5 &   \\ 
30 & 2FGL J0842.9$-$4721  &     &  08:42:58  &  $-$47:21:53  & 7.4& 7.1 & 11.4 & SPP \\ 
\\
31 & 2FGL J1814.1$-$1735c  &  1FGL J1814.0$-$1736c  &  18:14:09  &  $-$17:35:31  & 4.9& 4.3 & 11.0 &   \\ 
32 & 2FGL J1636.3$-$4740c  &  1FGL J1636.4$-$4737c  &  16:36:22  &  $-$47:40:58  & 4.3& 3.3 & 10.9 &   \\ 
33 & 2FGL J2022.8$+$3843c  &     &  20:22:50  &  $+$38:43:21  & 8.2& 7.7 & 10.6 & SNR \\ 
34 & 2FGL J1714.5$-$3829  &  1FGL J1714.5$-$3830c  &  17:14:31  &  $-$38:29:32  & 2.8& 2.3 & 10.5 & SPP \\ 
35 & 2FGL J1056.0$-$5853  &     &  10:56:00  &  $-$58:53:16  & 8.4& 7.2 & 10.5 &   \\ 
36 & 2FGL J1911.0$+$0905  &  1FGL J1910.9$+$0906c  &  19:11:03  &  $+$09:05:38  & 1.6& 1.5 & 10.4 & SNR \\ 
37 & 2FGL J1638.0$-$4703c  &     &  16:38:03  &  $-$47:03:10  & 3.9& 3.2 & 10.3 &   \\ 
38 & 2FGL J1536.4$-$4949  &  1FGL J1536.5$-$4949  &  15:36:30  &  $-$49:49:45  & 1.7& 1.6 & 10.2 &   \\ 
39 & 2FGL J1628.1$-$4857c  &     &  16:28:11  &  $-$48:57:36  & 11.8& 6.7 & 10.1 & SPP \\ 
40 & 2FGL J1311.7$-$3429  &  1FGL J1311.7$-$3429  &  13:11:46  &  $-$34:29:19  & 2.1& 2.0 & 10.1 &   \\ 
\\
41 & 2FGL J1650.6$-$4603c  &  1FGL J1651.5$-$4602c  &  16:50:36  &  $-$46:03:16  & 3.5& 3.1 & 10.1 &   \\ 
42 & 2FGL J1112.1$-$6040  &  1FGL J1112.1$-$6041c  &  11:12:07  &  $-$60:40:17  & 2.1& 2.0 & 9.9 & SPP \\ 
43 & 2FGL J0608.3$+$2037  &  1FGL J0608.3$+$2038c  &  06:08:20  &  $+$20:37:55  & 7.0& 6.4 & 9.9 &   \\ 
44 & 2FGL J1615.0$-$5051  &     &  16:15:02  &  $-$50:51:06  & 5.4& 4.6 & 9.9 & SPP \\ 
45 & \dag2FGL J1740.4$-$3054c  &  1FGL J1740.3$-$3053c  &  17:40:25  &  $-$30:54:41  & 10.1& 6.2 & 9.8 & SPP \\ 
46 & 2FGL J0340.5$+$5307  &  1FGL J0341.5$+$5304  &  03:40:36  &  $+$53:07:52  & 9.0& 7.6 & 9.7 &   \\ 
47 & 2FGL J2033.6$+$3927  &  1FGL J2032.8$+$3928  &  20:33:39  &  $+$39:27:05  & 7.2& 5.6 & 9.7 &   \\ 
48 & 2FGL J1914.0$+$1436  &     &  19:14:05  &  $+$14:36:15  & 9.6& 7.8 & 9.6 &   \\ 
49 & 2FGL J1027.4$-$5730c  &     &  10:27:27  &  $-$57:30:39  & 5.4& 4.7 & 9.6 &   \\ 
50 & 2FGL J1843.7$-$0312c  &     &  18:43:43  &  $-$03:12:56  & 8.1& 5.8 & 9.5 &   \\ 
	\hline \hline \end{tabular} 
\end{center} \end{table*}

\section{Conclusion and Discussion} \label{sec:con}

In this paper, we reviewed the Gaussian mixture model and its application
in data modelling and classification. As examples, we apply it to
pulsar classification in the $P-\dot{P}$ diagram as well as modelling
and ranking the 2FGL catalog point sources.

In the application of the GMM to model pulsar populations, we find that
the pulsar distribution in the \PPD diagram should be described by six
Gaussian clusters. Based on the GMM, we present a rule to separate the MSP
population from normal pulsar population.  As caveats, the six Gaussian
clusters from GMM algorithm may be artifacts due to the requirement
of approximating a non-Gaussian distribution of pulsars. Such a
non-Gaussian distribution can be induced by selection effects, pulsar
evolution, star formation history, etc. However these six Gaussian
components coincide with other observational evidence, e.g. chemical
composition of companion, radiation and polarization properties, timing
behaviors etc. Although it is still far from drawing any solid conclusion,
those clusters found by the GMM algorithm may imply: i) There are two
different MSP populations with different evolution scenarios, which is
supported by the evidence that one cluster contains mainly CO-white dwarf
companions and the other contains mainly He-white dwarfs companions. ii)
There are two possible different groups of normal pulsars. iii) High magnetic 
field
pulsars may come from the evolution of normal pulsars. iv) Although there are 
different formation channels, recycled pulsars appear to form a continuum in the 
\PPD diagram.  Hence the spectrum of accreted
mass for both fully recycled and mildly recycled MSPs should be smooth,
otherwise we would identify more clusters.

In the application to the 2FGL catalog, we use a 3-D parameter space spanned
by $V\!I$, $Sc$, and \FTH.  We found that the distribution can be well
described by three Gaussian clusters, two of which correspond to pulsar
and AG populations. The remaining cluster contains sources with lower
detection significance. Using the GMM, we calculate the pulsar likelihood
for each source and sort the 2FGL catalog accordingly. In the sorted
list, we find that the top 5\% sources contain 50\% known pulsars, the
top 50\% contain 99\% known pulsars, and no known active galaxy appears
in the top 6\%.  Clearly this algorithm has the ability to prioritize
the follow-up searching observation scheme to find new pulsars. The
statistical behavior of the sorted list is given in \FIG{fig:prob} and
a sample with the first and the last 5 sources of the sorted list is
presented in \TAB{tab:samp}. In \TAB{tab:cand}, we also present a list
of un-associated 2FGL catalog sources with high pulsar likelihood for future
pulsar searching projects.

We include the gamma-ray flux as one dimension of our parameter space,
although most discrimination comes from $V\!I$ and $Sc$. The reason to do
so is that both $V\!I$ and $S\!c$ correlate with the Test Statistics. By
introducing the gamma-ray flux, we can, to a certain degree, correct such
a correlation. Instead of Flux1000 (the integral flux from 1 to 100 GeV),
we have also tried using other flux measurements available in the 2FGL
catalog. In particular, we have experimented using Energy\_Flux100 (the
energy flux from 100 MeV to 100 GeV obtained by spectral fitting), which
yields a ranking result with a slightly lower detection rate compared to
the result using Flux1000. We have tried with four dimensional GMMs. In
these experiments, various photon flux, energy flux, Test Statistics of
different energy bands, and Galactic coordinates were tried as the forth
dimension. The classification results show little differences. This is
mainly due to the fact that most other variables available in the
2FGL catalog are correlated with the Flux1000, and thus do not provide much
extra information to improve the classification.

There are alternatives to the un-supervised machine learning techniques,
where one uses supervised machine learning \citep{Fermi11}. One particular
benefit of being un-supervised is that, since we use no prior information
of source associations in our ranking algorithm, we can investigate the
statistics quality of the algorithm easily by comparing the results with
known pulsar and AG populations, as shown in \FIG{fig:prob}.

We note that the GMM algorithm detects three clusters in the
\FTH-$V\!I$-$Sc$ parameter space, one of which is in the confused region
with low gamma-ray fluxes. Such component is likely to be an artifact due
to the low signal-to-noise statistic. 

\section{Acknowledgement} We gratefully acknowledge support from ERC
Advanced Grant ``LEAP'', Grant Agreement Number 227947 (PI Michael
Kramer). Y. L. Yue is supported by the National Natural Science Foundation
of China (grant 11103045), the National Basic Research Program of
China (grant 2012CB821800) and the Young Researcher Grant of National
Astronomical Observatories, Chinese Academy of Sciences. We thank Ramesh
Karuppusamy and Thomas Tauris for reading the manuscript and their very
helpful comments.  We also appreciate the valuable inputs and help from
the anonymous referee.

\label{lastpage}

\end{document}